\documentstyle[aps,twocolumn]{revtex}
\begin{document}
\draft

\title{Controlling atom-atom interaction at ultralow
temperatures by dc electric fields}
\author{M. Marinescu and L. You}
\address{School of Physics, Georgia Institute of Technology,
Atlanta, GA 30332-0430}
\date{\today}
\maketitle

\begin{abstract}
We propose a physical mechanism for tuning the
atom-atom interaction strength at ultra-low temperatures.
In the presence of a dc electric field
the interatomic potential is changed due to the effective
dipole-dipole interaction between the polarized atoms.
Detailed multi-channel scattering
calculations reveal features never before
discussed for ultra-cold atomic collisions.
We demonstrate that optimal control
of the effective atom-atom interactions can be achieved under
reasonable laboratory conditions. Implications of this research on
the physics of atomic Bose-Einstein condensation (BEC) and on the
pursuit for atomic degenerate fermion gases will be discussed.
\end{abstract}


\narrowtext

The study of weakly interacting quantum gases has attracted significant
attention since the initial success of Bose-Einstein condensation (BEC)
\cite{bec}.
Tremendous progress has been made over the last three years in both
theory and experiment. One of the recent interests is the study of
controlling the strength of atom-atom interaction. Several groups have
discussed mechanisms for
changing the scattering length of the atom-atom interaction
using near resonant lasers \cite{gora}, radio frequency  fields \cite{verhaar},
and Feschbach resonances induced by a magnetic field \cite{eite}.
Indeed, very recently, Feschbach resonances have been observed by
several experimental groups \cite{fre}.

This letter concerns the physics of adjusting the
effective low energy atomic interactions.
We propose the use of an external dc electric field (dc-E) to influence the
low energy atomic collisions by means of modifying the shape of the
interaction potential between atoms. This letter is organized as follows:
We start with a brief discussion of
the effective interaction potential between two alkali-metal atoms
in the presence of a dc-E, followed by
an outline of the main results of multi-channel collision
formalism for two polarized atoms at ultralow temperatures.
Illustrative results are presented, using a
typical model potential for the atom-atom interaction.
For collisions between bosonic atoms, we show that:
(i) the sign of the scattering length $a_{\rm sc}$
may be determined by measuring the relative deviation of the
total elastic cross section in the presence of a dc-E,
(ii) it is possible to tune the value of $a_{\rm sc}$ smoothly in a broad
range, (iii) a dc-E can induce strong anisotropic interactions
between the atoms, and (iv) a dc-E can induce zero energy
resonances (these are {\it shape resonances} in contrast to
{\it Feschbach resonances} recently observed \cite{fre}).
For collisions between fermionic atoms we
show that (v) the dc-E can induce strong interactions between
atoms in spin symmetrized states even at zero temperature.
We believe these results open the door for a new area
of studies of quantum degenerate atomic gases with adjustable
and anisotropic interactions.

In the usual treatment of the binary interaction between two
spherically symmetric atoms in the ground state, the long-range
interaction potential is given [in the London-van der Waals
(LvW) formalism] by the following expression
\begin{equation}
V(R)=-\frac{C_6}{R^6}-\frac{C_8}{R^8}-\frac{C_{10}}{R^{10}}-\cdots,
\label{a1}
\end{equation}
where $C_6$, $C_8$, and $C_{10}$ are the dispersion coefficients, and
$R$ is the internuclear distance. This is a ``short-range"
potential; therefore,
the zero energy scattering is described
essentially only by the S-wave scattering length, $a_{\rm sc}$.

In the presence of a dc-E the spherical
symmetry of the interacting atoms is distorted and consequently
the long-range form of the interatomic potential [ Eq. (\ref{a1})]
needs to be reconsidered. In the infinite separation limit,
the ground state wave function of the atoms in a dc-E
acquires a small $P$ angular momentum component along the electric field,
given by the dipole coupling between the $S$ ground state and the
$P$ excited states of the unperturbed atom.
To the leading order in the dc-E intensity $\cal E$,
the LvW formalism generates an additional term,
\begin{equation}
V_E(R)=-\frac{C_E}{R^3} P_2(\cos\theta),
\label{a2}
\end{equation}
where $C_E=2{\cal E}^2\alpha_1^A(0)\alpha_1^{B}(0)$ is the
electric induced dipole interaction coefficient and $\alpha_1^{A(B)}(0)$ are
the static atomic dipole polarizabilities of atom A and B respectively.
$P_2(.)$ is the Legendre polynomial of order 2 and $\theta$ is the angle
between the directions of the electric field and the internuclear axis.

We note that
in a complete treatment of the long-range interactions, other
terms induced by the dc-E are also present. The
first order perturbation (in terms of the Coulomb interactions
between the atomic charge distributions) generates an additional term
proportional to $1/R^5$,
and which is related to the quadrupole couplings between
the $P$ state components of the perturbed ground state of the
atoms in a dc-E. This term is proportional
to ${\cal E}^4$ and can be neglected for the ${\cal E}$ values
of interest to us in this study (which are very weak in
atomic units). The second order perturbation provides
corrections to the dispersion coefficients $C_6$, $C_8$, and $C_{10}$
from Eq. (\ref{a1}), which are proportional to ${\cal E}^2$ (and
higher powers of $\cal E$). These corrections are also neglected in
the present analyses, again assuming a weak dc-E.
Although quantitatively the values of the electric induced dipole
term from Eq. (\ref{a2}) is small ({\it e.g.} 100kV/cm is
equivalent to $1.94401\times10^{-5}$a.u.), it provides a
qualitatively different asymptotic behavior for the
interaction potential ({\it i.e.} $1/R^3$) with significant implications
for the scattering at very low energies.
The complete long-range interatomic potential is therefore
\begin{equation}
V({\vec R})=V_0(R)+V_E(R),
\label{a3}
\end{equation}
where $V_0(R)$ is the usual long-range dispersion form Eq. (\ref{a1})
in the absence of dc-E.

The potential from Eq. (\ref{a3}) is anisotropic, and requires a
multi-channel treatment of the scattering.
The T-matrix elements, $T_{lm}^{l'm'}$,
can be extracted from the asymptotic conditions imposed on the
partial wave channels. The total elastic cross section is give by
\begin{equation}
\sigma_{B(F)}=8\pi\sum_{l,l'={\rm even(odd)}}\sum_{m,m'}
|t_{lm}^{l'm'}|^2.
\label{a4}
\end{equation}
for bosons (B) and fermions (F) respectively, where
$t_{lm}^{l'm'}\equiv T_{lm}^{l'm'}/k$ are the reduced T-matrix
elements.

The main analytical result obtained is this: the
reduced T-matrix elements $t_{lm}^{l'm'}$ are finite quantities in the limit
of zero energy. This result may be intuitively understood
as follows. If a potential approaches zero as $1/R^n$ when $R$ goes
to infinity, then the phase shift $\delta_l$ behaves in the limit
of zero energy as $k^{2l+1}$ if $l<(n-3)/2$ and as $k^{n-2}$ otherwise.
Our problem is more complex since it involves a system of coupled equations.
However, the coupling terms are essentially $\sim1/R^3$.
Thus, the effective potential generated by the couplings, for each channel,
behaves in the limit of large $R$ as $1/R^6$. Then, the character
of the total effective potential for each partial wave equation is decided
by the diagonal part of the potential. For $l=0$
the effective potential behaves as $1/R^6$ and so $\delta_0\sim k$.
For $l\neq0$ the effective potentials behave as $1/R^3$  [generated by
the diagonal part of $V(E)$]
and so $\delta_l\sim k$. From these assertions we infer
that in our case all $t_{lm}^{l'm'}$ are finite
quantities in the limit of zero energy. A more rigorous proof
of these assertions will be presented elsewhere. Thus, the electric field
induced part of the potential, Eq. (\ref{a2}), has a ``quasi long-range"
character in the sense that it generates a ``short-range" contribution
to the effective potential of the partial wave channel $l=0$ while
it generates a ``long-range" contribution (proportional to $1/R^3$)
for all other partial wave channels.

To illustrate our results we present numerical results
for a model interaction potential.
This model contains all the features present in a real potential
curve and it has been studied in the context of
scattering length computations \cite{mm}. Moreover, by slightly changing the
cutoff radius $R_c$ \cite{mm}, this model generates a broad range
of values for the scattering length. We use it to simulate
qualitatively the results which may be obtained for actual
ground state potential curves of alkali-metal dimers.

Figure 1 presents the  logarithm of $\sigma_B$ in a.u.
as a function of the electric field
for six different potentials (defined by six different $R_c$ values:
23.226, 23.171, 23.155, 23.1446, 23.138, and 23.1245).
The curves are labeled by  the value of the
corresponding scattering length in the absence of the external field.
Since $\sigma_B$ has a very weak dependency on the
collision energy in the domain of sub-mK temperatures (regardless
of the value of the electric field), these results
may be interpreted as values
at zero collision energy. Note that curves corresponding to potentials
with a small scattering length contain a
resonance peak at certain values of the electric field. The smaller
the value of the scattering length, the smaller
the value of the electric field where the resonance occurs.
We note that for small values of the electric field
$\sigma_B$ tends to become
smaller for potentials with a positive scattering length and to
become larger for potentials with a negative scattering length.
This behavior may be explained by the fact that the
dc-E tends to lower the effective
interatomic potential curves, eventually allowing the
transformation of a virtual state into a new bound state.
During this process the value of $t_{00}^{00}$ will increase
with the electric field, which in turn yields a decrease of $\sigma_B$
for potentials with an initial positive scattering length
(negative $-t_{00}^{00}$) and an
increase of $\sigma_B$ for potentials with an
initial negative scattering length (positive $-t_{00}^{00}$).
Thus, by measuring the relative deviation of the
$\sigma_B$ in the presence of a dc-E,
one may determine the sign of the
scattering length. Given the extreme sensitivity of the dependence of
$\sigma_B$ on the interatomic potential
as well as on the value of the electric field, one may be able
to refine interatomic potentials by
analyzing the values of the $\sigma_B$ measured at different electric
field intensities. One way to investigate such properties is to
study the macroscopic effects of a dc electric field on
degenerate atomic gases at ultralow temperatures.

Figure 2 presents $\sigma_B$ in units of $8\pi a_{\rm sc}^2$, {\it i.e.}
$\sigma_b\equiv\sigma_B({\cal E})/\sigma_B(0)$ (solid line)
and the asymmetry parameter $\delta\equiv8\pi|t_{00}^{00}|^2/\sigma_B$
(dashed line) as a function of the
electric field for the potential with scattering
length $a_{\rm sc}=2470$ a.u.
The asymmetry parameter $\delta$ has a value close to 1 when $\sigma_B$
is dominated by the S-wave contribution and has a value close to zero
when $\sigma_B$ is dominated by other partial wave contributions.
In this case $\sigma_b$
decreases smoothly with the increase of the electric field.
Moreover, $\delta$ has values close to 1 in the region of maximum
variation of $\sigma_B$. Thus the effective scattering length,
$a_{\rm eff}\equiv-t_{00}^{00}$ is the only parameter needed to
characterize the collision process.
This fact suggests that for cases where the scattering length
has a large positive value one may obtain a smooth variation of the
$\sigma_B$ at zero temperature by varying the
external electric field. We envisage
interesting applications in the study of atomic BEC by controlling the
strength of the atomic interaction.

Figure 3 presents the values of the asymmetry parameter
$\delta$ as a function
of the electric field for the case of $a_{\rm sc}=32$ a.u.
The zero of $\delta$ is associated
with the zero of $t_{00}^{00}$ when
it changes sign from minus to plus,
which in turn is the result of the balance between the repulsive
effect of the last bound state and the attractive effect of the
first virtual state. This occurs because when the last bound state
is closer to the threshold
than the first virtual state the sign of $a_{\rm sc}$ is positive
and when the first virtual state is closer to the threshold than the
last bound state the sign of $a_{\rm sc}$ is negative.
In this region the
scattering process has a strong anisotropic character. This result
suggests that, by properly choosing the atomic species (with a moderate
positive value for the scattering length) and the value of the
electric field, one may study in a controlled fashion two
particular interesting regimes of the atomic BEC. One is the atomic
BEC with anisotropic atomic interactions and the other is the
study of the collapse of the BEC when the value of the effective
scattering length is changed from positive to negative.

In Fig. 4 we present
the values of the reduced T-matrix element $t_{00}^{00}$ (solid line)
as a function of the electric field
for the case of $a_{\rm sc}=-2121$ a.u.
As the figure suggests, $t_{00}^{00}$ has
an initial positive value and is increasing
with the electric field until a new bound state is supported by the
molecular system ({\it i.e.} a virtual state has been transformed into
a bound state) and a jump to $-\infty$ occurs.
Moreover, the asymmetry parameter $\delta$ (dashed line) has a value
close to 1 in the region where the resonance occurs, which indicates
that this resonance is an S-wave resonance. This result suggests
that by properly choosing the value of the electric field one may study
in a controlled fashion the zero energy resonances and their
influence on the atomic BEC. Also it appears possible to
obtain stable condensates for atomic species with a
negative scattering length
by tuning the effective scattering length to positive values.

Figure 5 presents the logarithm of $\sigma_F$ in a.u.
for two colliding fermionic atoms as a function of the
electric field. We note that the values of $\sigma_F$
are also insensitive to the collision energy in the domain of
sub-mK temperatures. In the fermionic case only the T-matrix
elements corresponding to odd partial wave channels for $l\geq1$
contribute to the final expression of $\sigma_F$.
Thus, the structure of the $\sigma_F$ from Fig. 5 is mainly
a result of competition between the centrifugal potential $l(l+1)/R^2$ and
the electric field induced part of the potential $V_E$, Eq. (\ref{a2}).
No major effects are expected to come from the small details of the
interatomic potential $V_0(R)$.
The numerical values
obtained for $\sigma_F$ are the same (in three digits) for all
potentials discussed in the bosonic case. We also note that the
increase in the value of $\sigma_F$ is significant.
For a potential with $a_{\rm sc}=32$ a.u., the zero energy $\sigma_B$ in the
absence of the external field is $2.574\times10^4$ a.u. This value
is obtained for $\sigma_F$ (see Fig. 5) at
298 kV/cm. Thus, the value of $\sigma_F$ in the
presence of a dc-E may be as large as the $\sigma_B$ for
an equivalent bosonic system in the absence of dc-E.
The fact that in the presence of a dc-E $\sigma_F$ is not zero
is a direct consequence of the modified collision asymptotes:
the reduced T-matrix elements are finite and non-zero quantities
in the limit of zero energy. This result strongly indicates
the possibility of studying interacting ultracold
atomic fermion gases in a controlled fashion.
The possibility of evaporative cooling of an atomic
fermi gas in the presence of a dc-E is also suggested.
It also opens the exciting new possibility of P-wave pairing
of magnetically trapped fermionic atoms.

In conclusion, we have studied atomic collisions
at ultralow temperatures in the presence of a dc electric field.
We found interesting low energy behaviors strongly encouraging new
directions of study for ultracold boson and fermion degenerate gases
with controlled atomic interactions.
We showed that for a bosonic system:
(i) study of the total elastic cross section for small values
of the electric field allows for determination of the sign of the
scattering length, (ii) the effective
scattering length may be smoothly changed in a broad range of values,
(iii) for certain values of the parameters, the atomic interaction
displays a strong anisotropic character, and (iv) in certain cases, the
zero energy resonances are accessible under realistic experimental
conditions. Similar possibilities exist for the control of the fermion
interactions
at ultralow temperatures. We have shown (v) that the interatomic
interaction strength may be boosted to significantly large values by an
external electric field.

This work is supported by the ONR grant 14-97-1-0633. L.Y. want
to thank Tony Leggett for interesting discussions about anisotropic
interactions.

\begin{figure}
\caption{The logarithm of the total elastic scattering cross section
for two colliding
boson-like atoms $\sigma_B$ as a function of the
electric field for six different interatomic
potentials. The curves are labeled by the value of the
scattering length in the absence of the external field.}
\end{figure}

\begin{figure}
\caption{Total elastic scattering cross section in units
of $8\pi a_{\rm sc}^2$, $\sigma_b$ (solid line), and
the asymmetry parameter $\delta$ (dashed line)
as a functions of the electric field for
the potential with a scattering length $a_{\rm sc}=2470$ a.u.}
\end{figure}

\begin{figure}
\caption{The values of the asymmetry parameter $\delta$ as a
function of the electric field for the potential with
a scattering length $a_{\rm sc}=32$ a.u.}
\end{figure}

\begin{figure}
\caption{The values of the $t_{00}^{00}$ reduced T-matrix element (solid line)
and the asymmetry parameter $\delta$ (dashed line)
as functions of the electric field for the potential with
a scattering length $a_{\rm sc}=-2121$ a.u.}
\end{figure}

\begin{figure}
\caption{The logarithm of the total elastic scattering cross
section $\sigma_F$ for two colliding
fermion-like atoms as a function of the electric field.}
\end{figure}

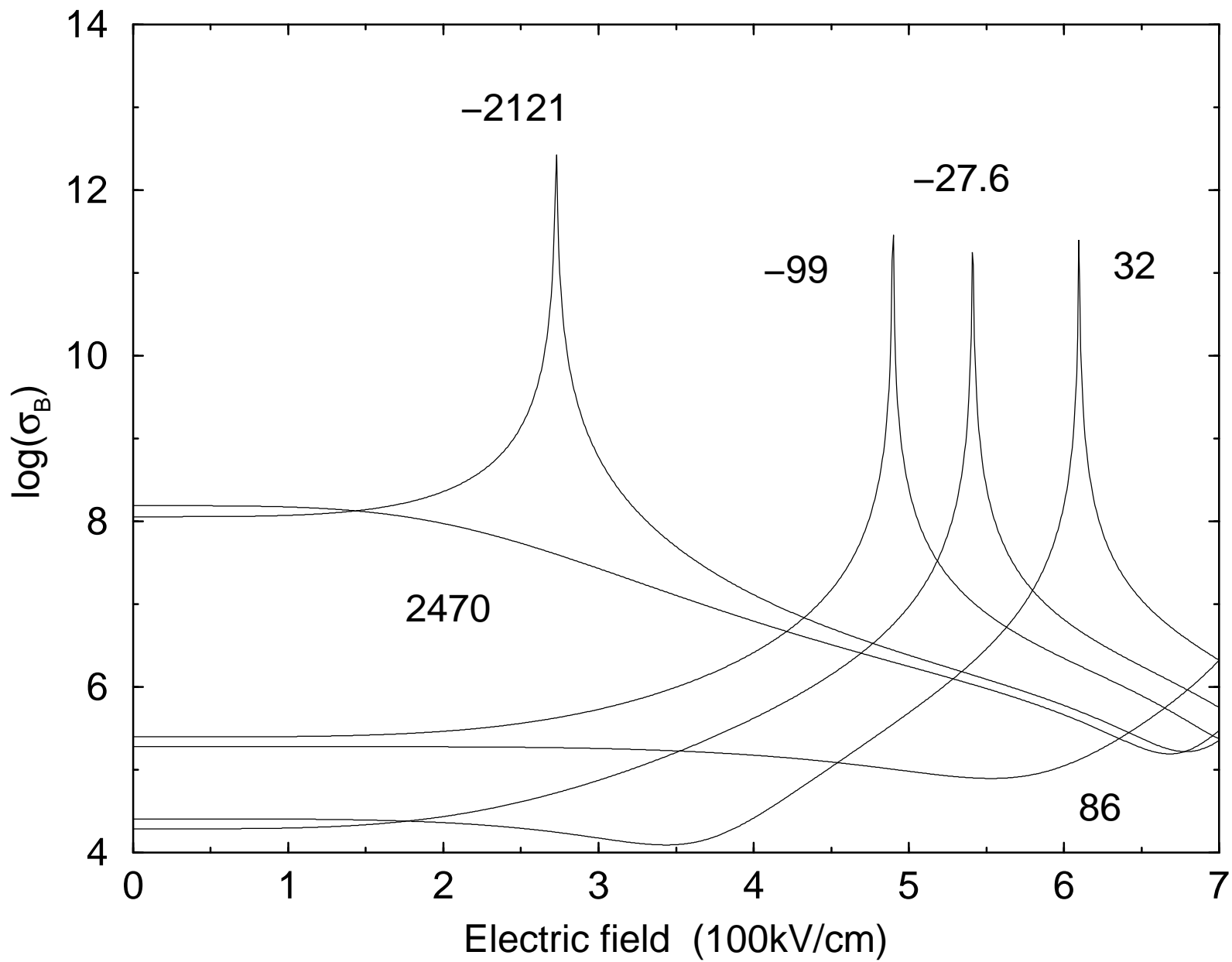
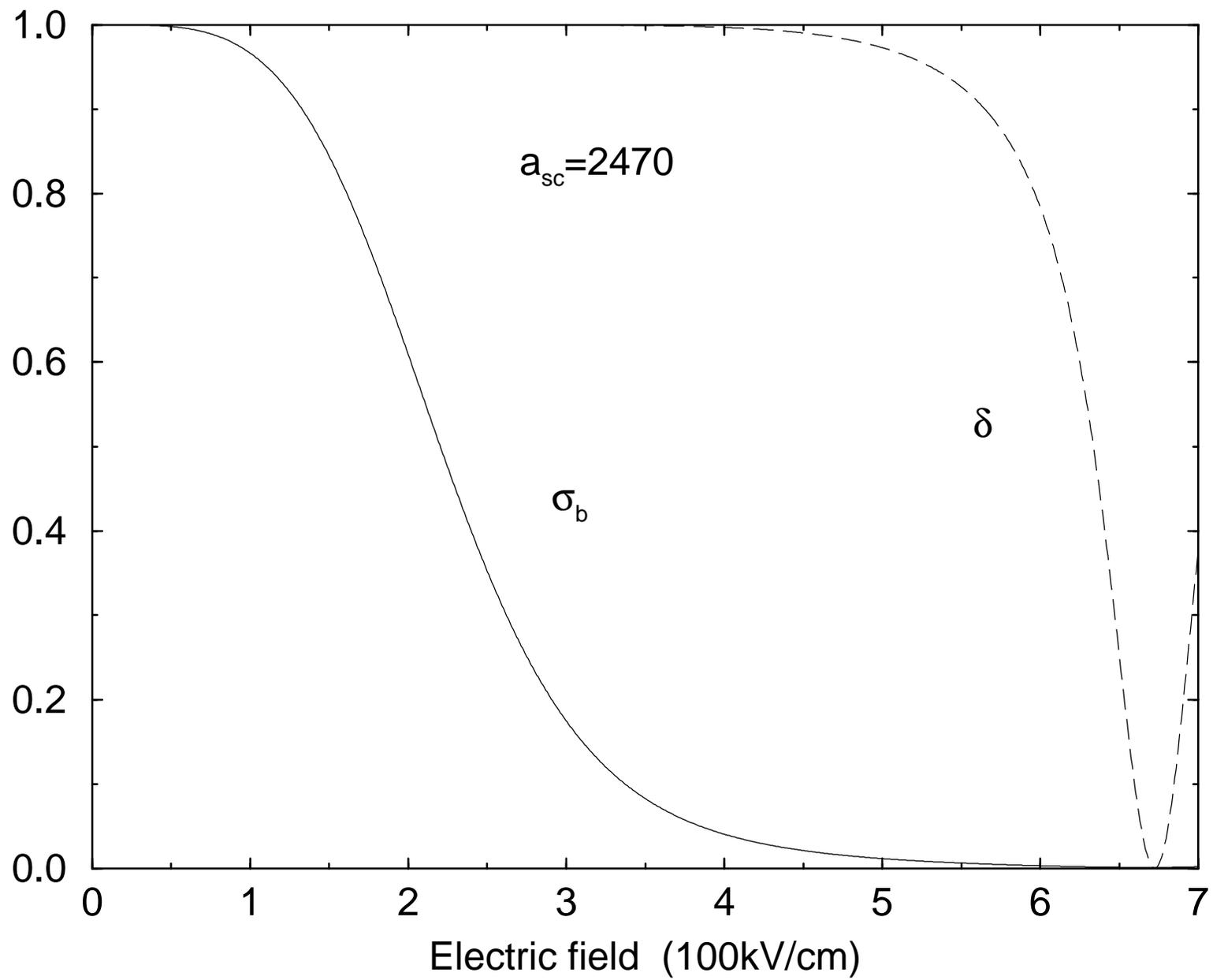
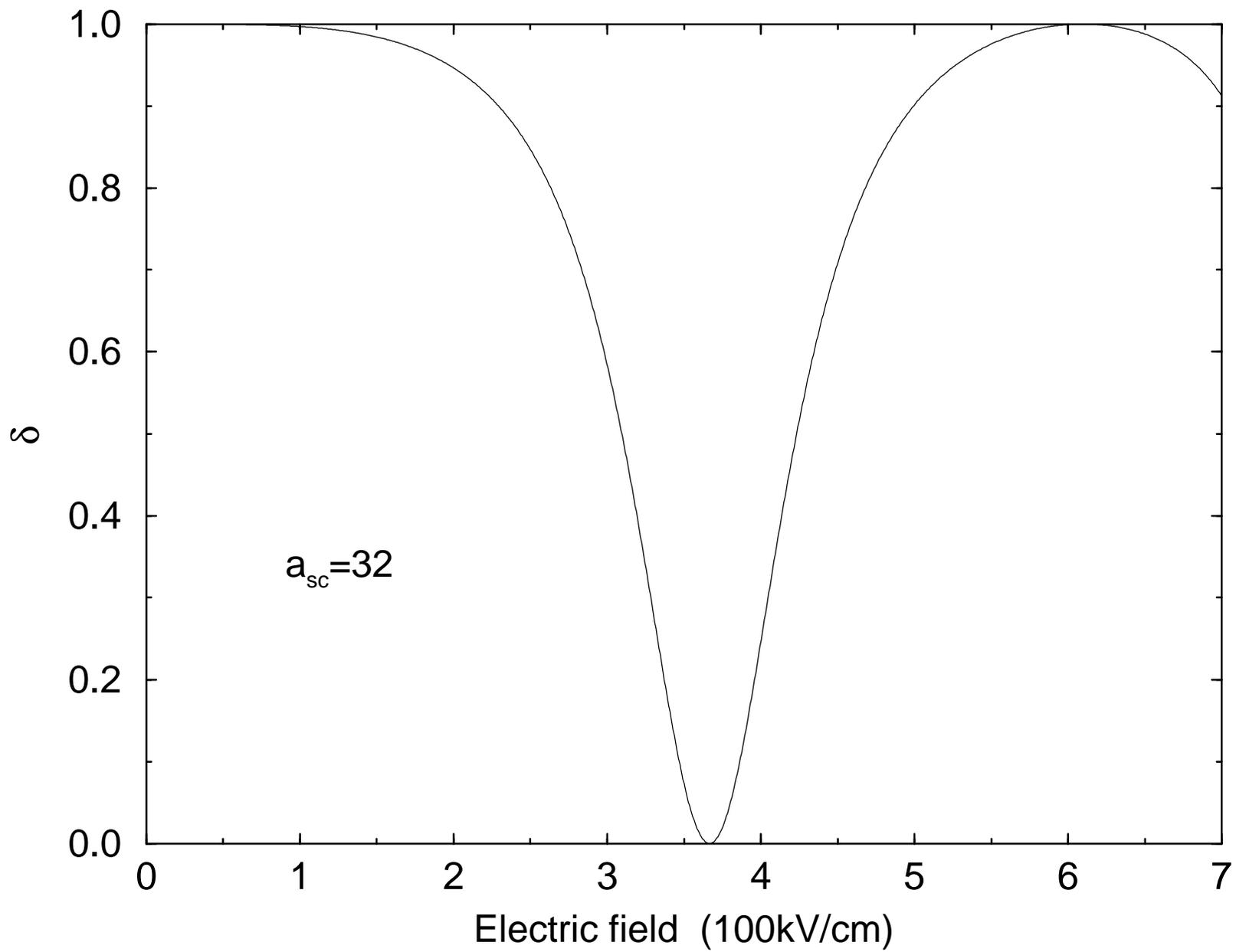
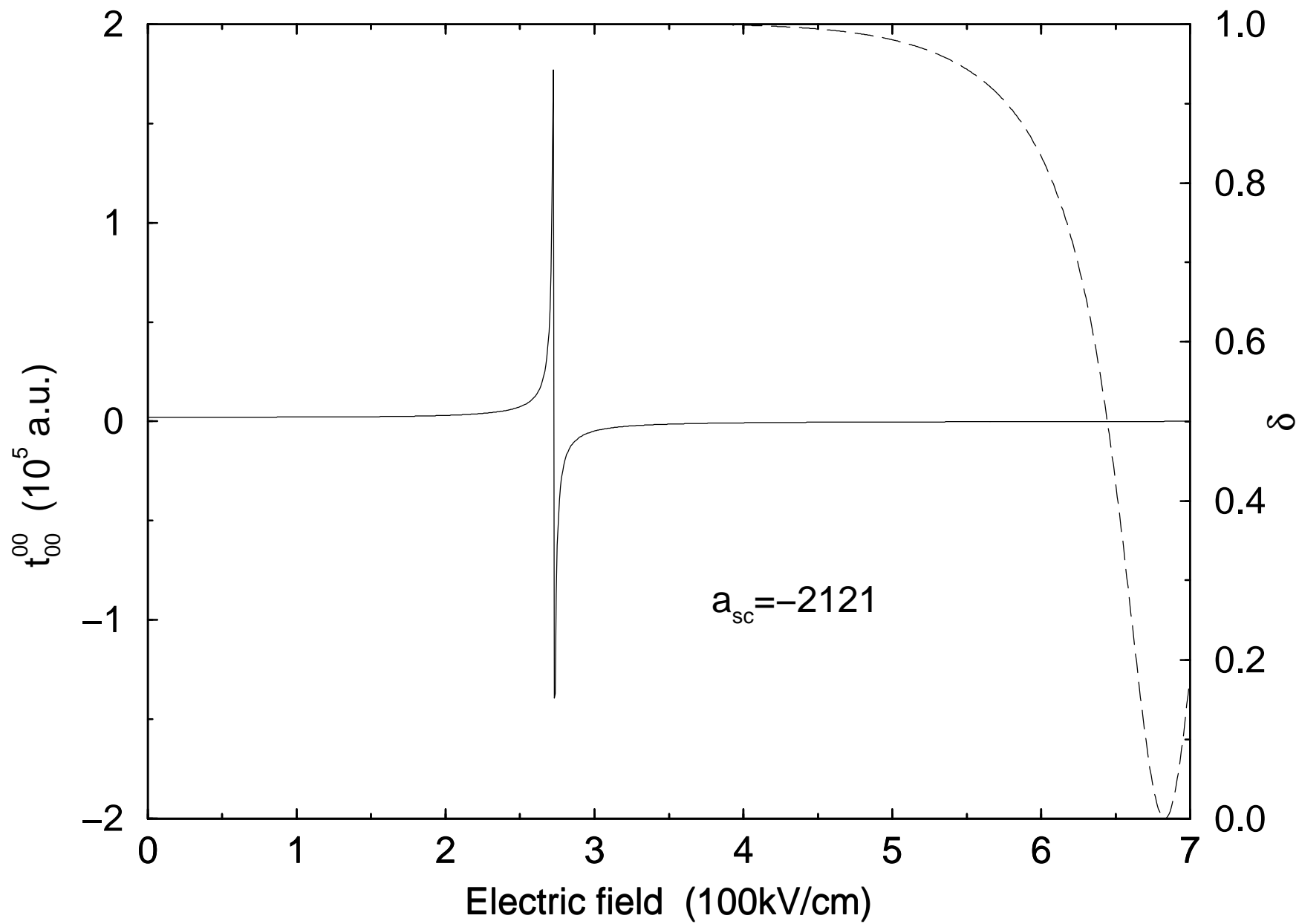
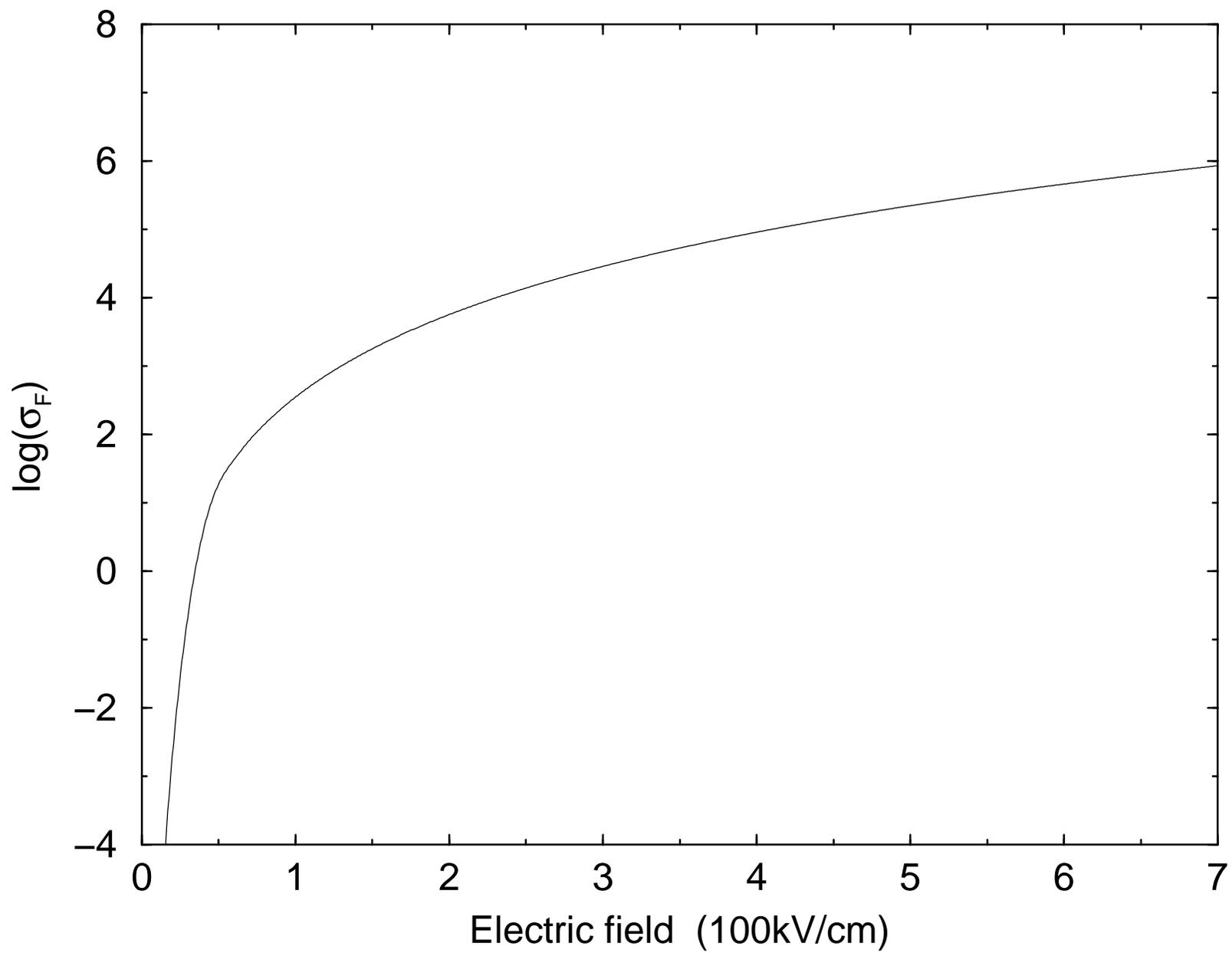
\end{document}